\documentclass[%
aps, pra,
 amsmath,amssymb,
 reprint,%
longbibliography, author-year]{revtex4-1}
\usepackage{amsmath}
\usepackage{graphicx}
\usepackage{dcolumn}
\usepackage{bm}
\usepackage[mathlines]{lineno}

\usepackage{breakurl}
\usepackage[breaklinks]{hyperref}

\usepackage[utf8]{inputenc}
\usepackage[T1]{fontenc}
\usepackage{mathptmx}
\usepackage{etoolbox}
\usepackage{CJKutf8}

\makeatletter
\def\@email#1#2{%
 \endgroup
 \patchcmd{\titleblock@produce}
  {\frontmatter@RRAPformat}
  {\frontmatter@RRAPformat{\produce@RRAP{*#1\href{mailto:#2}{#2}}}\frontmatter@RRAPformat}
  {}{}
}%

\usepackage{xcolor}
\hypersetup{
    colorlinks,
    linkcolor={red!50!black},
    citecolor={blue!50!black},
    urlcolor={blue!80!black}
}

\usepackage[english]{datetime2}
\DTMnewdatestyle{dashdate}{%
}
\DTMsetdatestyle{iso}

\usepackage[autostyle, english = american]{csquotes}
\MakeOuterQuote{"}

\usepackage{amsthm}

\def\bb{\mathbb}
\def\bf{\mathbf}

\def\N{\bb{N}}

\makeatother

\begin{document}
\begin{CJK*}{UTF8}{gbsn}


\title[]{Perfect diffusion is \texorpdfstring{$\mathsf{TC}^0$}{TC\^0} -- Bad diffusion is Turing-complete}

\author{Yuxi Liu} \affiliation{Berkeley Artificial Intelligence Research Lab, UC Berkeley}
\email{yuxi\_liu@berkeley.edu} \homepage{https://yuxi-liu-wired.github.io/}

\date{\today}

\begin{abstract}
This paper explores the computational complexity of diffusion-based language modeling. We prove a dichotomy based on the quality of the score-matching network in a diffusion model. In one direction, a network that \textit{exactly} computes the score function of some initial distribution can only perform language modeling within the $\mathsf{TC}^0$ complexity class, reflecting limitations tied to rapid convergence. In the other direction, we show that if there is no requirement for the network to match any score function, then diffusion modeling can simulate any Turing machine in a certain sense. This dichotomy provides a theoretical lens on the capabilities and limitations of diffusion models, particularly concerning tasks requiring sequential computation. We conjecture extensions of our theoretical results, including for the case where the diffusion model is not perfect, but merely good. We also discuss the wider context and practical implications, and hypothesize that a machine learning architecture that can interpolate between sequential and parallel modes of operation would be superior to both Transformers and diffusion models.
\end{abstract}

\maketitle
\end{CJK*}
 
\section{Introduction}

Diffusion modeling is a standard technique in generative modeling of probability distributions. First proposed in 2015 \cite{sohl2015deep}, it has been considered state of the art in generative modeling of images since 2021 \cite{dhariwal2021diffusion}. Researchers have recently extended these models beyond images to handle discrete states, including language modeling \cite{austin2021structured, gong2022diffuseq}. This expansion raises fundamental questions about the computational nature of diffusion processes that our paper aims to address.

A diffusion model can be used to sample from a distribution with a variable amount of computing steps. This is usually understood as an advantage, in the sense of providing a compute-precision tradeoff: With a few steps, one can sample from the distribution approximately, and with increasing number of steps, the distribution can be sampled from with increasing precision. However, this intuitive picture also suggest that this advantage may be a curse. Specifically, it suggests that after a few sampling steps, further computation is "wasted" in the sense that they refine the result in a way that does not matter, because the result has converged.

Indeed, empirically, diffusion models typically converge rapidly with a fixed number of denoising steps regardless of input. For example, Ravishankar et al. \cite{ravishankar2024scaling} studied using diffusion models for depth-perception, and showed that there is no difference between 5 and 100 sampling steps, Similarly, Austin et al. \cite{austin2021structured} showed that the perplexity of diffusion language modeling did not differ between 10 and 1024 diffusion steps.

While such rapid convergence may be regarded as an advantage, this also raises concerns about their fundamental capabilities. A key insight from language modeling research is that autoregressive models benefit from "chain of thought" approaches when tackling complex reasoning tasks. These approaches allow models to work through problems step by step before arriving at a final answer, much like human reasoning. This benefit isn't just observed in practice but has theoretical foundations \cite{feng2023towards, li2024chain}. In the big picture, both theory and practice has shown that chain of thought approaches benefit precisely in allowing a variable amount of sequential processing, which is necessary for solving certain kinds of "inherently sequential" problems.

As an example, if your program solve both easy and difficult Sudoku puzzles in exactly the same number of steps, you might reasonably question whether your method actually works for the hard puzzles. At one end of the limit, an easy Sudoku puzzle could be solved in a few steps by filling in each blank in parallel, since each blank could be solved by checking its row, column, and square. There is no sequential dependence between the blanks. At the other end of the limit, a hard Sudoku puzzle would involve a large amount of dependence between many blanks, which would require deep tree searches and significant backtracking to solve. It would be unreasonable to expect a parallel algorithm to solve a hard Sudoku puzzle, no matter the "width" of the parallelism, if it does not have the requisite "depth".

Indeed, certain kinds of empirical failures of diffusion language models suggest that they struggle precisely on tasks that require sequential processing. For instance, when using diffusion models to solve Sudoku puzzles, Wewer et al. \cite{wewer2025spatial} found that denoising all digits simultaneously worked for easy puzzles but failed for difficult ones. Performance improved only when denoising fewer digits at a time, with optimal results achieved by denoising just one digit at a time, essentially reverting to a purely sequential process.

Similarly, Arriola et al. \cite{arriola2025block} observed that discrete diffusion models underperform compared to autoregressive approaches. The solution they proposed simply reintroduces autoregressive generation, applying diffusion to generate a few tokens at a time. This effectively recreates the autoregressive language modeling paradigm, with diffusion being merely a stand-in replacement for the usual Transformer decoder.

These empirical observations suggest a deeper theoretical question: Is there a fundamental computational limitation to diffusion models that prevents them from solving certain problems efficiently? In this paper, we provide a theoretical framework in which we can formalize and prove certain stylized facts from these empirical findings. In one direction, we prove that "perfect" diffusion models are constrained to the $\mathsf{TC}^0$ complexity class due to their rapid convergence. In the other direction, we explicitly construct "bad" ones to perform any Turing-computable operation, precisely because they do not converge. We make certain conjectures and extensions for further work.

\section{Setup}

Before we begin presenting the theorem and the construction, we set up the mathematical framework.

\subsection{Diffusion modeling}

Diffusion models work by gradually adding noise to data and then learning to reverse this process. Think of it as watching a drop of ink spread through water, and then learning to recover the original drop from the diluted state. This intuitive physical analogy connects to their mathematical foundation, which borrows concepts from thermodynamics and statistical physics.

In machine learning literature, there are two main formulations to describe this process. The first, \textbf{Denoising Diffusion Probabilistic Models (DDPM)}, approaches the problem through discrete time steps. The second, \textbf{score-matching with Langevin dynamics (SMLD)}, uses continuous differential equations. Despite their different origins, these two approaches are fundamentally equivalent, as demonstrated by \cite{kingma2021variational} and \cite{luo2022understanding}.

The connection works in both directions. DDPM can be seen as a discretized version of SMLD, where each DDPM update step corresponds to using the Euler–Maruyama method to solve SMLD's stochastic differential equation (SDE). Conversely, if we take the limit of infinitely many DDPM steps with infinitesimally small noise additions, we recover the continuous SDE formulation of SMLD. This equivalence means that models trained using either framework can be used interchangeably for sampling purposes.

For clarity and mathematical convenience, we primarily use the SMLD formulation throughout this paper, though our results apply equally to both frameworks, since they are equivalent.

Consider a\textbf{ data distribution} $\rho_{data}$ over the real space $\mathbb{R}^d$. The task of SMLD is to learn a \textbf{score-matching function} $f_\theta$ that allows us to sample from $\rho_{data}$.

A \textbf{noise schedule} is a continuous function $\beta$ of type $[0, \infty) \to [0, \infty)$, such that $\beta(t)$ can be interpreted as the noising rate in the forward diffusion process at time $t$. We require $\int_0^\infty \beta(t) dt = \infty$, which can be interpreted as saying that eventually all signal is destroyed, leaving only noise. 

Define the distribution at $t=0$ by $\rho_0 := \rho_{data}$. Suppose we sample a data point $x_0 \sim \rho_0$, and let it evolve according to the SDE
\begin{equation}
dx_t = -\frac{1}{2} \beta(t) x_t dt + \sqrt{\beta(t)} dW_t
\end{equation}
then this implies a time-evolution of the data distribution over time, which can be directly solved by the Fokker-Planck equation:
\begin{equation}
\partial_t \rho_t = \frac{1}{2} \beta(t) (\nabla \cdot (x \rho_t) + \Delta \rho_t)
\end{equation}

At the $t \to \infty$ limit, the distribution converges to the standard normal distribution $\mathcal{N}(0, I_d)$.

For any time $T > 0$, the time-evolution can be exactly reversed as follows. Let $\hat{x}_T$ be sampled according to $\rho_{\hat{x}, 0} := \rho_T$, then the following SDE equation would lead to an exact reversal:
\begin{equation}
d\hat{x}_t = \frac{1}{2} \beta(T-t) \hat{x}_{t} dt + \beta(T-t) \underbrace{\nabla_{\hat{x}_{t}} \ln \rho_{T-t}(\hat{x}_{t})}_{\text{score function}} dt + \sqrt{\beta(T-t)} dW_t
\end{equation}
where by "reversal" we mean that $\rho_{\hat{x}, t} = \rho_{T-t}$ for any $t \in [0, T]$ \cite{anderson1982reverse}.

Assuming that a score-matching function $f_\theta$ has been trained, such that 
\begin{equation}
f_\theta(x, t) \approx \nabla_x \ln \rho_t(x)
\end{equation}
for all $t, x$, then $\rho_{data}$ can be approximately sampled by initializing a pure-noise sample $\hat{x}_T \sim \mathcal{N}(0, I_d)$, then solving the backward SDE
\begin{equation}
\hat{x}_{t-dt} = \frac{1}{2} \beta(t) \hat{x}_t dt + \beta(t) f_\theta(\hat{x}_t, t) dt + \sqrt{\beta(t)} dW_t
\end{equation}
by any SDE integration method, such as Euler–Maruyama method. By varying the sizes of the $dt$ steps in the Euler–Maruyama method, we can recover different noise-schedules for DDPM.

If $f_\theta(x, t) = \nabla_x \ln \rho_t(x)$ is exact, then at the limit of $T \to \infty$ and infinitely many steps in the Euler–Maruyama method, we can exactly sample from $\rho_{data}$.

\subsection{Circuit complexity}

A circuit complexity class is a particular form of computational complexity class. In our paper, we focus on the $\mathsf{TC}^0$ class, which is particularly suited to studying the computational complexity of neural networks, because a family of feedforward neural networks with a constant number of layers is essentially a $\mathsf{TC}^0$ circuit family. Indeed, the class of $\mathsf{TC}^0$ were first proposed specifically in the 1980s to model the computational complexity of neural networks. \cite{parberry1988parallel}

Formally, $\mathsf{TC}^0$ is defined as the class of problems that can be decided by a family of boolean circuits with the following properties:

\begin{itemize}
\item \textbf{Boolean circuits:} A boolean circuit is a directed acyclic graph where each node (or gate) computes a boolean function of its inputs. The inputs to the circuit are boolean variables, and the output is a single boolean value.

\item \textbf{Unbounded fan-in:} Each gate in the circuit can receive inputs from an arbitrary number of other gates. This contrasts with bounded fan-in circuits where gates have a limited number of inputs. Convolutional neurons have bounded fan-in, but fully-connected neurons have unbounded fan-in.

\item \textbf{Polynomial width:} The number of gates at each level of the circuit is bounded by a polynomial in the input size $n$.

\item \textbf{Constant depth:} The longest path from any input to the output is bounded by a constant that does not depend on the input size. This may be interpreted as stating the circuit family is "highly parallelizable".

\item \textbf{Threshold gates:} A threshold gate is a binary neuron. It can be written as a function $\theta(\sum_i w_i x_i + t)$, where $w_i, t$ are real numbers, and $\theta$ is the binary step-function
\end{itemize}

For those unfamiliar, here is a short exercise: With 1 layer, we can construct "gadgets" such as the AND gate, the NOT gate, and all other common boolean gates with threshold gates. With 2 layers, we can construct the $k$-EQUALS gate for each $k$, which outputs 1 if exactly $k$ inputs are 1, and 0 otherwise. With 3 layers, we can construct the IS-IN gate for any finite subset of $\N$.

From the definition, it is clear that each member $\mathsf{TC}^0$ circuit family is essentially a feedforward neural network. However, this only consists of a single member. Since the neural network has a fixed number of inputs, it would be unable to process more inputs than the number of neurons in its input. This brings the idea of a circuit \textit{family}. A circuit family is a set of circuits $C_1, C_2, \dots$ such that each $C_n$ is capable of processing exactly inputs of length $n$. Computational complexity theory studies not the complexity of problems solvable by a single circuit, but a circuit family, because any single circuit is merely equivalent to a lookup table, and the complexity of the problem it solves is trivial.

Consequently, a $\mathsf{TC}^0$ family of feedforward neural networks is defined as a set of neural networks $C_n$, such that there exists a constant $D$ (the upper bound on depth), and a polynomial $p$, such that each $C_n$ has depth $\leq D$ and number of neurons $\leq p(n)$.

While the $\mathsf{TC}^0$ class is most similar to feedforward fully-connected neural networks, this is not necessarily the case. Indeed, a family of bounded-depth polynomial-width Transformers is still in the $\mathsf{TC}^0$ class. \cite{li2024chain}

\subsection{Language modeling}

At the most abstract level, a language is simply a set of words made of letters. Formally:

\begin{itemize}
    \item An \textbf{alphabet} $\Sigma$ is a finite nonempty set. Each element in the alphabet may be called a \textbf{letter} or a \textbf{token}.
    \item A \textbf{word} in an alphabet $\Sigma$ is a finite sequence of elements of $\Sigma$.
    \item A \textbf{language} $L$ in an alphabet $\Sigma$ is a set of words in the alphabet $\Sigma$.
\end{itemize}

A \textbf{prefix language modeling problem} is, given a sequence of tokens $x_1, \dots, x_n$, to compute the next token $x_{n+1}$. This is a deterministic formalization of "next-token prediction", the dominant paradigm in language modeling since the GPT-2 of 2019.

An example would be the word problem for finite groups: Given a finite group $G$, and a sequence of elements in the group $g_1, \dots, g_n$, compute $\prod_{i=1}^n g_i$. Intuitively, there is a method that computes this in $\log_2(n)$ parallel steps by binary multiplication: the first parallel step computes $g_1g_2, g_3g_4, \dots$, and so on. Since $\log_2(n)$ is not constant, this would not lie within the $\mathsf{TC}^0$ class. For certain groups, there are shortcuts to this process. For example, for any prime number $p$, the word problem in the mod-$p$ multiplicative group is computable in constant number of parallel steps via Fermat's little theorem. However, shortcuts probably do not exist in general. Indeed, if $G$ is the permutation group on 5 elements, then the corresponding word problem is not in the $\mathsf{TC}^0$ class, assuming widely believed conjectures in computational complexity theory. \cite{liu2023transformers}

While usually, a diffusion model is used for generating from a continuous state space such as $\mathbb{R}^d$, it can be used to model discrete distributions as well. This is necessary for language modeling. We consider the case closest to continuous state space modeling -- quantization: One divides the continuous state space $\mathbb{R}^d$ into regions, and assigns a token to each region. This then allows sampling a discrete distribution from a diffusion model with continuous state space. Formally, if $\Sigma = \{a_1, a_2, \dots, a_M\}$ is the alphabet, then we divide $\mathbb{R}^d$ into $M$ regions $V_1, \dots, V_M$, such that each region $V_i$ maps to a token $a_i$.

Also, as usual in circuit complexity theory, we need more than a single score-network $f_\theta$, but rather, a full sequence of them, so we define a \textbf{$\mathsf{TC}^0$ family of score-networks} to be a family of feed-forward neural networks $f_{\theta, 0}, f_{\theta, 1}, \dots$ , such that:
\begin{itemize}
\item Each $f_{\theta, n}$ takes as input $n+2$ elements $x_1, \dots, x_n, x, t$, and produces an output $f_{\theta, n}(x, t | x_1, \dots, x_n)$.
\item The family $f_{\theta, n}$ has $O(1)$ depth and $\mathsf{poly}(n)$ width.
\end{itemize} 

\textbf{Comment.} For the theorem to hold, it is not necessary to assume the family of neural networks are feed-forward. The theorem holds for any family of score-networks for which a single forward pass is in $\mathsf{TC}^0$. This includes, for example, Transformers and state-space models \cite{merrill2023parallelism, merrill2024illusion}.

Finally, since a diffusion model may solve a problem only with high enough probability, instead of solving it deterministically, we make the following definition: A prefix language modeling problem is \textbf{solved with constant probability bound} if there exists some $\epsilon > 0$, such that for each input token sequence $x_1, \dots, x_n$, let $x_{correct}$ be the correct response, then 
\begin{equation}
p(x_{correct}|x_1, \dots, x_n) > p(x'|x_1, \dots, x_n) + \epsilon, \quad \forall x' \neq x_{correct}.
\end{equation}

\subsection{Counter machines}

To show that a deliberately bad diffusion model may be Turing-complete, we show how they could simulate a particular kind of Turing-complete abstract machines: the counter machines. This is not necessary for understanding the theorem on "perfect" diffusion models.

A \textbf{counter machine} can be thought of as \textbf{finite-state automata} augmented with memories, each of which can hold a single unbounded integer. In our paper, we use the following form of counter machine, lightly modified from \cite{fischer1968counter}:

\begin{itemize}
\item The machine has access to a finite number $k$ of \textbf{registers}, notated as $r_1, \dots, r_k$. Each register stores a single integer.
\item The machine also has access to a read-only \textbf{input tape}, on which the machine has a read-head that can be moved in either direction. At machine start-up, the input tape has contents $\verb!^!a_1a_2\dots a_n\verb!$!$, where $\verb!^!$ and $\verb!$!$ denote the beginning and the end of the word, and $n$ is the length of the input word. The read-head is placed at the character just after $\verb!^!$, which may be  $\verb!$!$ if the input word is empty.
\item A \textbf{program} for the machine is a numbered list of instructions.
\item Each \textbf{instruction} is of the following format: conditional on the state of the read-head on the input tape and on whether each register is zero or not, modify every register by an amount in $\{-1, 0, +1\}$, move the read-head by up to one position in either direction, then jump to another instruction.
\item There is a special instruction named "HALT". If the machine arrives at such an instruction, it halts. Each HALT instruction may be marked as either an \textbf{accepting} HALT, or a \textbf{rejecting} HALT.
\item To \textbf{accept} an input word means the machine reaches an accepting HALT state. Similarly for \textbf{rejection}.
\item A \textbf{decider} for a language is a machine that accepts words in the language, and rejects words out of the language. It must halt on all inputs.
\end{itemize}

It is known that counter machines are Turing-complete, in the sense that a universal Turing machine can be simulated by a counter machine with 2 registers \cite{minskyComputationFiniteInfinite1967}. This implies in particular that any language that is decidable by a Turing machine is decidable by a counter machine.

\section{Main theorem}

\textbf{Theorem.} Suppose there exists a $\mathsf{TC}^0$ family of score-networks $f_{\theta, 0}, f_{\theta, 1}, \dots$, such that for each $n$ and each $x_1, \dots, x_n$, the function $f_{\theta, n}(x, t | x_1, \dots, x_n)$ exactly computes the score function of \textit{some} initial distribution $\rho_{0, n}$ with bounded first moment: $\mathbb{E}_{x_0 \sim \rho_{0, n}}[\|x_0\|] \leq 1$.

If this family solves a prefix language modeling problem at the limit of infinite time SMLD with constant probability bound, then the problem is in the $\mathsf{TC}^0$ class.

\textbf{Proof.} The idea of the proof is simple. We first construct a universal $O(1)$-bound on how many steps are sufficient for sampling the SMLD within a constant probability bound, then we derandomize it according to \cite{hajnal1993threshold}, while still remaining within the $\mathsf{TC}^0$ class.

Since the score-network exactly computes the score function for $\rho_0$, the score-matching error is exactly zero.

By \cite{li2024d}, there exists a constant $c > 0$ that does not depend on any parameter in the statement of the theorem (a "universal constant"), such that there exists a noise schedule for DDPM that takes $T$ steps of the following kind. Using the schedule and the score-matching function $f_\theta$, we sample from a distribution $\rho_{DDPM, T}$. It satisfies the inequality 
\begin{equation}
TV(\rho_{DDPM, T}, \rho_{\hat{x}, 0}) \leq c \frac{d (\log T)^3}{T}
\end{equation}

Now, since $c$ is a universal constant, we can take $T$ to be $cc' d^2$, to obtain an upper bound
\begin{equation}
TV(\rho_{DDPM, T}, \rho_{\hat{x}, 0}) \leq \frac{(\log(cc' d^2))^3}{c'd}
\end{equation}
The key is that $T$ does not increase with $n$.

Since the growth of $\log^3$ is dominated by linear growth, for any $\epsilon' > 0$, there exists a large enough universal constant $c'$, such that
\begin{equation}
TV(\rho_{DDPM, T}, \rho_{\hat{x}, 0}) \leq \epsilon'
\end{equation}
for all $d = 1, 2, 3, \dots$.

Let $\epsilon$ be the constant probability bound, then setting $\epsilon' = \epsilon / 2$, we find that $\rho_{DDPM, T}$ already solves the problem with constant probability bound.

Now we can derandomize this family, obtaining a $\mathsf{TC}^0$ family of boolean circuits that solves the problem deterministically. The details of the derandomization method appears in \cite[Proposition 4.2]{hajnal1993threshold}. The big picture is as follows: we remove the probability by hard-coding a magic constant (also known as a "non-uniform advice string") per member of the family, such that sampling for polynomially many times, and taking the majority output, would always give the correct output. By Hoeffding's inequality, such magic constants exists for a large enough polynomial. $\blacksquare$

\textbf{Comment.} The requirement for \textit{exact} score-matching is necessary for the following two reasons:

First, the full form of the inequality from \cite{li2024d} is
\begin{equation}
TV(\rho_{DDPM, T}, \rho_{\hat{x}, 0}) \leq c \frac{d (\log T)^3}{T} + c \epsilon_{\text{score}} \sqrt{\log T}.
\end{equation}
where the term $\epsilon_{\text{score}}$ denotes the score-matching error between the true score function of $\rho_{\hat{x}, 0}$ and the approximation $f_\theta$. As this extra term \textit{increases} with $T$, the proof above does not apply.

Second, if we have \textit{no} requirement on score-matching, then there is essentially no constraint on the computational power of SMLD, by the following construction.

Practically relevant score-networks are intermediate between two extreme cases. We believe that if $f_\theta$ is a good enough, but not perfect, score-matching network, then a generalized version of the above theorem still applies. However, finding the right way to \textit{quantify} the goodness, as well as proving such a generalization, is left as future work.

\section{Turing-computability construction}

As previously stated, the counter machine formalism is an alternative formalism for computability, equivalent to the Turing machine formalism. Given a counter machine with $k$ registers, we describe how to construct a "pinball" machine that operates according to the SDE
\begin{equation}
d\hat{x}_t = \frac{1}{2} \hat{x}_t dt + f_\theta(\hat{x}_t, t) dt + dW_t
\end{equation}
under a smooth force field $f_\theta$, thus showing Turing-completeness.

The pinball machine has a single ball, whose location is $\hat{x}_t$. The ball moves around a state space $\mathbb{R}^d$ guided by the force field $f_\theta(\hat{x}_t, t)$. Indeed, the force field can be time-independent, so we write it as $f_\theta(\hat{x}_t)$ instead.

We divide the state space $\mathbb{R}^d$ into three parts as $\mathbb{R}^d = \mathbb{R}^k \times \mathbb{R} \times \mathbb{R}$. The first part $\mathbb{R}^k$ represents the registers. The second part $\mathbb{R}$ represents the program counter, which tracks the line-number of the program. The third part $\mathbb{R}$ is used for jumping between instructions, providing enough room for maneuver without "crossing the wires".

The space is divided into cubic cells of side lengths $L$. We denote each cell by $k + 2$ integers. 

Corresponding to the tripartite structure of the state space, the force field has three parts as well. One part simply cancels out the $\frac{1}{2} x_t$ term. Another part behaves like grooves, along which the ball rolls, thus implementing the counter machine. The third part points towards the center-lines of the grooves, so that the ball is not knocked off the grooves by the noise term $dW_t$.

Instead of formally specifying the grooves, it is simpler to give an example. Suppose at line number 32, the instruction reads "If the current state of register 1 is zero, then increment register 2 and jump to line 23, else jump to line 33", then this is implemented by drawing the following paths:
\begin{itemize}
\item $(0, r_2, \dots, r_d, 32, 0) \to (0, r_2 + 1, \dots, r_d, 32, 32) \to (0, r_2 + 1, \dots, r_d, 23, 32) \to (0, r_2 + 1, \dots, r_d, 23, 0)$.
\item $(r_1, r_2, \dots, r_d, 32, 0) \to (0, r_2, \dots, r_d, 32, 32) \to (0, r_2, \dots, r_d, 33, 32) \to (0, r_2, \dots, r_d, 33, 0)$ for nonzero $r_1$.
\end{itemize}

By smoothing the corners of the paths, we obtain a smooth force field. $\blacksquare$

\textbf{Comment.} We can more carefully handle the problem of leakage: The noise term $dW_t$ would, over a long enough time, eventually knock the ball off the grooves. This can be suppressed by either using a strong confinement force field, or by using a weak confinement force field but a large cubic cell side length $L$. Specifically, we show how to obtain a confinement force field with Lipschitz-continuity coefficient bounded by a universal constant.

Let the counter machine have $N$ instructions. Suppose it halts within $S$ steps, then the total distance travelled by the pinball would be $O(NSL)$, where we need to account for the time necessary to jump between instructions. Then, since the rate of leakage is on the order of $e^{-L^2}$, we need only require $L \geq O(\sqrt{\ln (NSL)})$ to suppress the probability of leakage during the entire computation to a small constant. In particular, for any fixed $N, S$, because $L$ grows faster than $\sqrt{\ln (NSL)}$, there exists a big enough $L$ for which the machine will halt without leakage, for probability as close to $1$ as one desires. This machine operates under a force field that is smooth, and has Lipschitz-continuity bounded by universal constant.

Suppose that we have a language that is decidable by a Turing machine when it is restricted to a working tape with length $O(f(n))$, where $n$ is the input length, and $f$ is some monotonically increasing function, then by \citep[Theorems 3.1 and 3.2]{fischer1968counter}, it is decidable by a counter machine that takes $e^{O(f(n))}$ steps to halt. Thus, it suffices when $L \geq O(\sqrt{f(n)})$.

\section{Future work}

\subsection{Theoretical extensions}

We have shown that perfect diffusion models with exact score matching are constrained to $\mathsf{TC}^0$, while deliberately "bad" diffusion models can be Turing-complete. The more realistic intermediate case remains open, where the score network approximately computes the score function.

We conjecture that similar computational limitations apply when the approximation quality is sufficiently high, but formalizing this notion of "sufficiently good approximation" and proving the corresponding result requires further work. We make this conjecture based on two reasons. One, the aforementioned empirical observation that diffusion models converge rapidly. Two, because the forward diffusion converges exponentially rapidly to the standard normal distribution $\mathcal N(0, I)$, we believe that the backward diffusion process, as long as it is sufficiently close to the score function of a forward diffusion process, would be forced to converge in $O(1)$ time, since exponential decay is fast decay.

Our analysis focuses on diffusion models operating on $\mathbb{R}^d$ with subsequent discretization. However, other formulations of discrete diffusion exist, such as the directly discrete approach in \cite{austin2021structured}. For these models, we conjecture that $\mathsf{TC}^0$ limitations apply regardless of score network quality, as the finite state space inherently constrains the ``computational capacity'' of the diffusion process. Intuitively, a finite state space allows encoding only a finite number of bits per state before the signal-to-noise ratio\footnote{See \cite{kingma2021variational} for a formalization of signal-to-noise in diffusion modeling.} is exhausted, and the reverse diffusion reaches $t=0$.

Between finite state spaces and $\mathbb{R}^d$ lies the intermediate case of continuous but compact state spaces, such as the unit ball in $\mathbb{R}^d$. While our ``pinball machine'' construction would still work in such spaces, it would require dividing the compact space into an increasing number of cells. This means the force field, while smooth, cannot maintain bounded Lipschitz-continuity coefficients. Because of this, we hypothesize that under the additional requirement of $O(1)$ Lipschitz-continuity, diffusion models on compact spaces would be constrained to $\mathsf{TC}^0$ regardless of score network quality, effectively making them computationally equivalent to finite state models.\footnote{A reviewer stated that the paper should end here. We somewhat agree, but the academic style requires the paper to continue. Feel free to stop reading here.}

\subsection{Empirical validation}

The animating big-picture idea behind this paper is that certain tasks are inherently sequential, such that any parallel computation that takes too little sequential steps must necessarily err. Sequential processing and consequences of its lack has been systematically studied for Transformers under the name of "chain of thought", but not for diffusion models. We have collected a few suggestive examples gleaned from the literature, but it would be a valuable contribution to the literature to test this hypothesis systematically on diffusion models. We conjecture:

\begin{itemize}
\item Tasks requiring deep sequential reasoning should exhibit a sharp performance cliff when addressed by diffusion models with a fixed number of denoising steps.
\item Adding more denoising steps beyond a certain threshold should yield minimal improvements for $\mathsf{TC}^0$ tasks but continued improvements for tasks outside this complexity class.
\item Performance on complex sequential tasks should improve significantly when introducing autoregressive components, as seen in \cite{arriola2025block}.
\end{itemize}

Controlled experiments testing these predictions would provide valuable empirical validation of our theoretical framework and guidance for the further development of diffusion models.

\subsection{Architecture}

Some architectures are inherently more sequential than others. Recurrent neural networks, for instance, process inputs step by step. They were state of the art in language modeling, before being largely replaced by Transformers because of the latter's massive parallelism, but sequential processing has made a comeback through chain-of-thought approaches when sequential reasoning proved necessary for complex tasks.

A similar pattern emerged in the recent history of image generation. The OpenAI DALL-E system used a standard Transformer architecture to generate images patch-by-patch. Diffusion models then took over with their fully parallel generation across all pixels. If our thesis about computational constraints holds, we might expect a return to some form of sequential inference for advanced image generation tasks.

The most promising direction may be architectures that interpolates sequential and parallel computation dynamically, shifting to the sequential mode for tasks that demand them. We point out several particularly worthy directions for interpolation:

\begin{itemize}
    \item In architecture, interpolation between massively parallel models (Transformers, state-space models) and sequential ones (recurrent neural networks).
    \item For language generation, interpolation between full-sequence generation (typical of diffusion language models) and autoregressive generation (common in Transformer-based models). While both approaches have been studied extensively in isolation, their combination remains relatively unexplored.
    \item Interpolation between SMLD and neural ODE frameworks. SMLD offers rapid convergence through massive parallelism, while neural ODEs provide slower convergence with more sequential computation.
\end{itemize}

\section{Conclusion}

This paper established a fundamental dichotomy in the computational capabilities of diffusion models for language modeling. We proved that "perfect" diffusion models, characterized by exact score matching to some initial distribution, are inherently limited. Their rapid convergence constrains their computational power to the $\mathsf{TC}^0$ complexity class. This theoretical limit aligns with empirical observations suggesting that standard diffusion models struggle with tasks demanding deep sequential reasoning, converging quickly regardless of problem complexity.

Conversely, we demonstrated that this limitation is not absolute for all diffusion-like processes. By constructing a "bad" diffusion model with a specifically designed, non-convergent score function, we showed that such systems can achieve Turing-completeness, capable of simulating arbitrary sequential computation via a pinball machine analogy.

These findings highlight a potential trade-off inherent in the diffusion framework: the efficiency gained through rapid, parallel denoising may come at the cost of computational depth required for inherently sequential tasks. While the precise complexity of realistic models with approximate score matching remains an open question, our results provide a theoretical grounding for understanding the observed strengths and weaknesses of diffusion models, suggesting that incorporating mechanisms for controlled sequential processing, as explored in some recent works, may be crucial for extending their capabilities to more complex reasoning domains.

\bibliographystyle{alpha}
\bibliography{references}

\end{document}